\documentclass[twocolumn, twocolappendix, numberedappendix]{openjournal}

\usepackage{natbib}

\usepackage{xcolor}
\definecolor{xlinkcolor}{cmyk}{1,1,0,0}
\usepackage[breaklinks,colorlinks,linkcolor=xlinkcolor,citecolor=xlinkcolor,urlcolor=xlinkcolor]{hyperref}

\usepackage[T1]{fontenc}

\usepackage{orcidlink}

\usepackage{xspace}

\usepackage{graphicx}

\usepackage{aas_macros}
\usepackage{booktabs}
\usepackage{longtable}
\usepackage{appendix}
\usepackage{amsmath}
\usepackage{amssymb}
\usepackage{graphicx}
\usepackage{fontawesome}

\newcommand{\persec}{s$^{-1}$}

\begin{document}

\title{\vspace{-0.8cm}Using the Circumgalactic Medium of Dwarf Galaxies as a \\ Calorimetric Dark Matter Detector\vspace{-1.5cm}}

\author{Abby Mintz\,\orcidlink{0000-0002-9816-9300}$^{1}$\footnote{abby.mintz@princeton.edu}, Digvijay Wadekar\,$^{2,3,4}$, Romain Teyssier\,$^{1}$
}
\affiliation{
$^{1}$Department of Astrophysical Sciences, Princeton University, 4 Ivy Lane, Princeton, NJ 08544, USA\\
$^{2}$Center for Gravitational Physics, University of Texas at Austin, Austin, TX 78712, USA\\
$^{3}$School of Natural Sciences, Institute for Advanced Study, Princeton, NJ 08540, USA\\
$^{4}$Department of Physics and Astronomy, Johns Hopkins University,
3400 N. Charles Street, Baltimore, Maryland, 21218, USA\\
}

\begin{abstract}
Many well-motivated dark matter (DM) scenarios, including dark photon DM, axion-like particles, sterile neutrinos, and primordial black holes, can deposit energy into ordinary matter and significantly raise the thermal energy of astrophysical gas. We present a simple, model-independent implementation of DM-induced heating in the \texttt{RAMSES} hydrodynamic simulation code, parameterized by an effective local heat-transfer rate $\Gamma_{\rm heat}$. Using simulations of an isolated gas-rich dwarf galaxy with properties similar to WLM, we find that DM heating substantially alters the thermal structure of the circumgalactic medium (CGM) while leaving the star formation rate and the dense ISM nearly unchanged. For $\Gamma_{\rm heat} = 10^{-27}$ to $10^{-25}\ {\rm s}^{-1}$, the neutral hydrogen column at 10~kpc is suppressed by 0.6 to 3.9~dex relative to the control simulation. Similarly, the \ion{C}{2} and \ion{Si}{2} column densities decrease by 4.1 and 5.5~dex, and by 2.2 and 3.0~dex for \ion{C}{4} and \ion{Si}{4}. These signatures are accessible to quasar absorption-line spectroscopy, so CGM observations of nearby field dwarf galaxies provide a promising avenue for constraining dark matter models that inject heat into ordinary matter.

\end{abstract}

\section{Introduction}

Astrophysical gas can act as a calorimeter for dark matter (DM) models that transfer energy to ordinary matter. If DM decay, annihilation, or scattering deposits heat into gas inside a halo, the resulting thermal response can be constrained by comparing the anomalous heating rate to the gas cooling rate \citep{DubHer15, WadFar21, WadWan22, WadWan23, BhoBramante18, BhoBra20, LuTak21, Tak21b, TakLu21, LahLu20, Kim20}. This idea is especially powerful in low-mass, gas-rich dwarf galaxies, where the gas density and metallicity are low and the astrophysical cooling rate is correspondingly small \citep{WadWan23, Sim19}.
Many non-standard DM scenarios can produce such heat exchange, including DM decay or annihilation to photons or $e^\pm$, axions and axion-like particles \citep{PQ,Raffelt:1990yz,Cad12,Marsh:2015xka}, sterile neutrinos \citep{Dodelson:1993je,Shi:1998km,Merle:2013gea,Boyarsky:2018tvu}, dark photon DM \citep{AnPosPra13_HPDM,AnPosPra15_HPDM,PigCleGui15_HPDM,WadFar21}, excited or inelastic DM states \citep{Fin07,Fin09,Bar20}, primordial black holes \citep{LuTak21, Tak21b, TakLu21, LahLu20, Kim20,WadWan23}, millicharged DM \citep{barkana18, barkana+18, WadFar21}, and other DM-baryon scattering scenarios \citep{Hainje:2026qou}.

The circumgalactic medium (CGM) is a particularly useful target for this test. The CGM is the gaseous envelope surrounding the galaxy disk but lying within the DM halo, and in isolated dwarfs it can survive without being removed by ram-pressure stripping \citep{Tumlinson2017_CGM}. Because dwarf galaxies have shallow potential wells, their CGM is sensitive to stellar feedback, photoionization, and other heating processes \citep{Bul17_Dwarf_Review, Collins2022}, all of which can strongly affect its composition. The ion content of the CGM can be traced using far-ultraviolet (FUV) spectroscopy of quasar sightlines passing through dwarf halos \citep{2014ApJ...796..136B, 2014MNRAS.445.2061L, Joh17_CGM_observations, 2018ApJ...864..132B, Zhe24_CGM_observations, Mis24_CGM_observations, 2026ApJ...996L..30J, Polzin2026}. FUV spectra contain numerous metal resonance lines that provide observational access to the gas temperature and ionization state.

Previous analytic calculations have shown that gas-rich isolated dwarfs can give strong constraints on DM heating \citep{WadWan22, WadWan23,WadFar21}. Those studies assumed that the gas is in hydrostatic equilibrium and required the anomalous heating rate not to exceed the astrophysical cooling rate. This calorimetric approach is complementary to conventional searches for photons, $e^\pm$, and CMB energy injection from DM decay or annihilation \citep{Ove04,Ess13,Sla16,Liu:2016cnk,BolChl20,Ali21,Gew17,Foster:2021ngm}. Here we go beyond the hydrostatic approximation in analytic calculations and perform hydrodynamic simulations, which self-consistently include ordinary astrophysical heating and cooling alongside DM heating. Related efforts have studied DM heating with hydrodynamic simulations \citep{Iwa17, Iwa19, Lis19, Lis20} and semi-analytic methods \citep{Nat07, Sch14, Sch18_DM_annihilation}. In particular, \citet{Lis19} implemented DM-annihilation heating in a particle-based hydrodynamics code and tested it in an isolated halo and a cosmological simulation, focusing on cosmological structure formation and comparatively massive systems. We instead target a low-mass dwarf galaxy, where the low gas density and metallicity reduce the cooling rate and make the response to anomalous heating especially pronounced. Our Eulerian mesh formulation also makes a local heating source straightforward to implement, avoiding the localized deposition and individual-timestep complications that arise in particle-based Lagrangian methods. Hydrodynamic simulations have also been used to distinguish the effects of other DM models, including warm DM \citep{Ros24} and atomic DM \citep{Roy2023_AtomicDM}. Recent RAMSES simulations have demonstrated the value of hydrodynamic modeling for interpreting the gas and stellar structure of low-mass dwarfs \citep{Blana2024_LeoT_RAMSES}. In this work we implement a model-independent DM heating source term in RAMSES \citep{Tey02_RAMSES} and apply it to an isolated dwarf galaxy simulation.

This paper is organized as follows. In Section~\ref{sec:numerical_methods}, we describe our RAMSES implementation of local DM heating, the isolated WLM-like dwarf galaxy simulations, and the post-processing used to predict neutral-hydrogen and metal-ion column densities. Appendix~\ref{app:timestep} discusses operator-splitting accuracy and possible timestep constraints for the heating source. In Section~\ref{sec:results}, we show how varying the model-independent heat-transfer rate $\Gamma_{\rm heat}$ changes the thermal and ionization state of the gas, with the largest effects appearing in the CGM. In Section~\ref{sec:discussion_conclusions}, we discuss how constraints on $\Gamma_{\rm heat}$ could be inferred from these signatures and mapped onto particular particle-physics models, and we describe limitations of the present isolated-galaxy simulations and future applications to dwarf galaxies formed from cosmological initial conditions.

\section{Numerical Methods}
\label{sec:numerical_methods}

\subsection{Dark Matter Heating Model}

We parameterize the local DM heating rate as
\begin{align}
    \frac{dE}{dt dV} &= \rho_{\rm{DM}}c^2\, \Gamma_{\rm{decay}}\frac{E_\mathrm{heat}}{E_\mathrm{particle}} \\
    &\equiv 9\times10^{-6} \frac{\rho_{\rm{DM}}}{1\, \rm{g/cm}^3}\frac{\Gamma_{\rm{heat}}}{10^{-26}\ \rm{s}^{-1}} \frac{\rm{erg}}{\rm{cm}^3\, \rm{s}}
\label{eq:DM_heating_rate}
\end{align}
where $E_\mathrm{heat}$ is the heat energy injected in the gas due to interactions of the decay products with the gas and $E_\mathrm{particle}$ is the rest mass energy for the non-relativistic DM particles that we consider. We therefore vary the effective, model-independent quantity $\Gamma_{\rm heat} \equiv \Gamma_{\rm decay} E_\mathrm{heat}/E_\mathrm{particle}$ directly in our simulations. For a particular DM model, the corresponding constraint on $\Gamma_{\rm decay}$ can be obtained if needed. For example, \citet{WadWan22} shows the explicit computation of the fraction of the decay energy deposited locally as heat ($E_\mathrm{heat}/E_\mathrm{particle}$) for models with DM decaying to photons such as axion-like particles and sterile neutrinos.

\subsection{Numerical Implementation in RAMSES}
\label{sec:ramses}

We use the adaptive-mesh-refinement code RAMSES \citep{Tey02_RAMSES} to evolve the gas, stars, and dark matter in an isolated dwarf galaxy. RAMSES solves the Euler equations on an adaptive Cartesian mesh and evolves collisionless particles with a particle-mesh gravity solver. The source term for DM heating introduced in this work is local: the heat deposited into a gas cell depends on the dark matter density assigned to that same cell (motivated by heat injection due to decay of DM particles). This approximation is appropriate for the model-independent heat-transfer parameterization used here, while non-local transport of decay products (see e.g., \cite{Liu:2019bbm}) is left to future work.

The implementation keeps the ordinary RAMSES Poisson density unchanged and adds a separate dark-matter-only density field, $\rho_{\rm DM}$. During the particle deposition step, RAMSES already assigns particle mass to the mesh for gravity. We use the same cloud-in-cell deposition scheme to construct $\rho_{\rm DM}$, but include only particles whose RAMSES particle family is classified as dark matter. Gas, stellar particles, sink/cloud particles, and tracer particles are therefore excluded from the heating source.

RAMSES uses operator splitting, i.e., it advances hydrodynamics, DM heating, and radiative cooling in successive substeps rather than simultaneously. On each refinement level, we add the DM heating source to every leaf cell after the hydrodynamic update and before cooling and chemistry:
\begin{equation}
    E_{\rm gas}^{n+1} = E_{\rm gas}^{n} + \rho_{\rm DM}\, c^2\,\Gamma_{\rm heat}\,\Delta t ,
\label{eq:update}
\end{equation}
where $E_{\rm gas}$ is the cell-centered total gas energy density and $\Delta t$ is the local RAMSES timestep. Cooling therefore sees the added heat immediately, whereas the hydrodynamic fluxes respond during the next hydrodynamic substep. Appendix~\ref{app:timestep} discusses the stability and accuracy of this split update. We make the patch implementing this DM-heating source term publicly available.\footnote{\label{fn:code}\url{https://github.com/JayWadekar/ramses-dm-heating}}

\subsection{Grid-Based Implementation and Timestepping}
\label{sec:timestep}

Adding a source term to the gas energy equation motivates a discussion of timestepping: the update must remain stable and sufficiently accurate when coupled to the hydrodynamic evolution. We will explain our approach and compare it with the particle-based DM-heating scheme introduced by \citet{Lis19} in this subsection. \citet{Lis19} implemented DM-annihilation heating in \texttt{GIZMO}, using its meshless finite-mass hydrodynamic solver to evolve moving gas elements in an approximately Lagrangian frame. In contrast, \texttt{RAMSES} evolves the gas on an Eulerian AMR grid. This difference makes a local DM-heating source simpler to implement in \texttt{RAMSES}. In the donor-based scheme of \citet{Lis19}, the energy carried by a single DM particle is deposited into a small number of neighboring gas elements. The resulting localized hot spot can drive a blast wave into the surrounding cold medium, motivating their timestep limiter based on Sedov--Taylor blast-wave propagation. In our implementation, cloud-in-cell deposition constructs a cell-centered DM density (Section~\ref{sec:ramses}), so the source is smooth on the cell scale and no point-like injection occurs.

The second reason particle-based implementations require special timestep machinery is that neighboring gas elements can be advanced at different times. A newly heated element may therefore interact with a cold neighbor that is still inactive, delaying the hydrodynamic response to the injected energy. In \texttt{RAMSES}, cells on the same refinement level advance together, and the recursive AMR update synchronizes neighboring levels. The grid therefore responds coherently to the deposited heat without an additional neighbor-timestep limiter.

In both particle- and grid-based implementations, decay- and annihilation-like sources add energy to the gas. If the deposited heating rate is held fixed during a timestep, the energy increment is proportional to $\Delta t$ and cannot make the internal energy negative. This makes implementing additional heating more straightforward compared to implementing cooling processes. Equation~\ref{eq:update} therefore integrates our decay-like source exactly for fixed $\rho_{\rm DM}$ and $\Gamma_{\rm heat}$, and the source update itself requires no stability limiter. The remaining concern is accuracy: because heating is operator-split from hydrodynamics and cooling, the timestep should be short enough to resolve the gas response. We discuss this point further in Appendix~\ref{app:timestep}. 

\subsection{Baryonic Physics}\label{subsec:simphysics}

Apart from the additional DM heating source term, we use a standard RAMSES galaxy-formation setup. Gas heating and cooling are implemented with equilibrium models for H and He \citep{Katz1996}, with additional metal cooling following \citet{Sutherland1993}. We include a uniform UV background \citep{Haardt1996}. Throughout this work, we assume \citet{Planck2020} cosmological parameters.

Star formation is modeled with a Schmidt relation where star formation is triggered in cells with density exceeding 100 H cm$^{-3}$. The star formation rate follows $\dot{\rho_\star} = \epsilon_\star \rho_g/t_\text{ff}$ where $\epsilon_\star$ is the star-formation efficiency, $\rho_g$ is the gas density, and $t_\text{ff}$ is the free-fall time. We fix $\epsilon_\star=0.01$. The simulation also includes feedback from supernovae, which inject mass, metals, and momentum into the surrounding cells. Each star particle returns a fraction $\eta_{\rm SN}=0.163$ of its initial mass as ejecta, releasing $10^{51}$~erg per 10~M$_\odot$ of ejecta at a constant mean rate between 3 and 20~Myr after formation. We assume the metal yield is 0.178.

\subsection{Initial Conditions, Resolution, and Simulation Suite}

To demonstrate the utility of our DM heating model, we run a set of simulations of an isolated dwarf galaxy constructed to resemble the general properties of the isolated Local Group galaxy Wolf--Lundmark--Melotte (WLM). Dwarf galaxies are useful targets because their lower gas densities and metallicities reduce the astrophysical cooling rate relative to larger galaxies (see Section~\ref{sec:Cooling_Rates}). The initial conditions were generated with the \texttt{DICE}\footnote{\url{https://bitbucket.org/vperret/dice/src/master/}} code based on parameters similar to those used in \citet{Andersson2023} and \citet{Hu2023}, which are in turn based on observationally derived values for WLM's structure and composition \citep[e.g.,][]{Leaman2012,Rubio2015,Mondal2018}. The DICE setup writes the particle initial conditions in Gadget2 format and includes distinct dark matter, stellar-disk, and gaseous-disk components.

The galaxy consists of a dark matter halo and a disk of gas and stars. The halo follows an NFW profile with concentration $c=15$, virial radius $R_{\rm 200c}\approx 46.5$ kpc, virial mass $M_\text{vir}\approx10^{10}$ M$_\odot$, spin parameter $\lambda=0.04$, and circular velocity $v_{200}=31.8$~km~s$^{-1}$ at the virial radius. The galaxy has a baryonic mass fraction of $0.8\%$, a stellar mass $M_{\star\text{, disk}} \approx 10^7$ M$_\odot$ consistent with estimates for WLM \citep{Lelli2016_SPARC}, and a gas mass $M_\text{g, disk} \approx 7\times10^7$ M$_\odot$ (i.e., a gas fraction $f_\text{gas} = 0.875$). The stellar and gaseous disks have identical exponential radial profiles with a scale length of 1.1 kpc and squared hyperbolic-secant vertical profiles with a scale height of 0.7 kpc:
$$\rho_{i,\text{disk}} = \rho_{i,0}\  e^{-r/r_{i,\text{disk}}}\cdot \text{sech}^2\frac{|z|}{z_{i,\text{disk}}} $$
where $i$ refers to either the gaseous or stellar component, $r_{i\text{, disk}}$ is the disk scale length, $z_{i,\text{,disk}}$ is the disk scale height, $r$ is the radial distance from the center of the disk, $z$ is the vertical distance from the disk midplane, and $\rho_{i,0}=M_\text{i,disk}/(4\pi r_{i\text{,disk}}^2 z_{i\text{,disk}})$ is the central density of the disk. The gas is initialized with a metallicity of 0.1\,Z$_\odot$, comparable to WLM's mean nebular oxygen abundance of approximately 13\% of the solar value \citep{Lee2005_WLM_metallicity}. The simulation domain is a $(60\,{\rm kpc})^3$ box. In addition to the DICE particle components, the gaseous halo is initialized with a low uniform density of $n_H =10^{-7}$ cm$^{-3}$, a uniform temperature of $T=10^5$ K, and a metallicity of 0.005 Z$_\odot$.

Our simulations are initiated with uniform grids with the coarsest resolution ($l_\text{min}=7$) of 128$^3$ covering the $(60\,\mathrm{kpc})^3$ box, corresponding to cells with side length $\sim470$pc. Cells are refined in a quasi-Lagrangian manner; if a cell contains baryonic mass exceeding 800 $M_\odot$ or more than 8 dark matter particles, refinement is triggered. The maximum resolution is set to $l_\text{max}=14$ for a maximum physical resolution of $\sim4$ pc.

We run four simulations with identical initial conditions and astrophysical physics, varying only the DM heating rate. The control simulation has $\Gamma_{\rm heat}=0$, while the three heating runs use $\Gamma_{\rm heat}=10^{-27}$, $10^{-26}$, and $10^{-25}\ {\rm s}^{-1}$. 

We run each simulation to $t=1.5$~Gyr, or slightly over one halo crossing time. This interval is long enough for the CGM to respond dynamically to the added heating. We do not integrate further because an isolated box has no steady state to approach: the open boundaries are not replenished by cosmological accretion, so gas driven beyond the box is permanently lost. By the end of the run, the simulation volume retains 98\% of its initial gas mass in the control run, but only 79\% at $\Gamma_{\rm heat}=10^{-26}$~\persec\ and 62\% at $10^{-25}$. The columns we quote are therefore not equilibrium predictions but the CGM response over a defined time interval. A real dwarf galaxy would have been exposed to DM heating for a Hubble time, but its CGM would also be continuously resupplied by cosmological accretion; the balance between these effects can only be determined with a cosmological simulation (Section~\ref{sec:cosmological_simulations}). The disk itself reaches a quasi-steady state well before the end of the run: in the control simulation, the disk neutral-hydrogen mass and star formation rate are constant
to within 2\% over the final 450~Myr.

\begin{figure}
\centering
\includegraphics[width=1\linewidth]{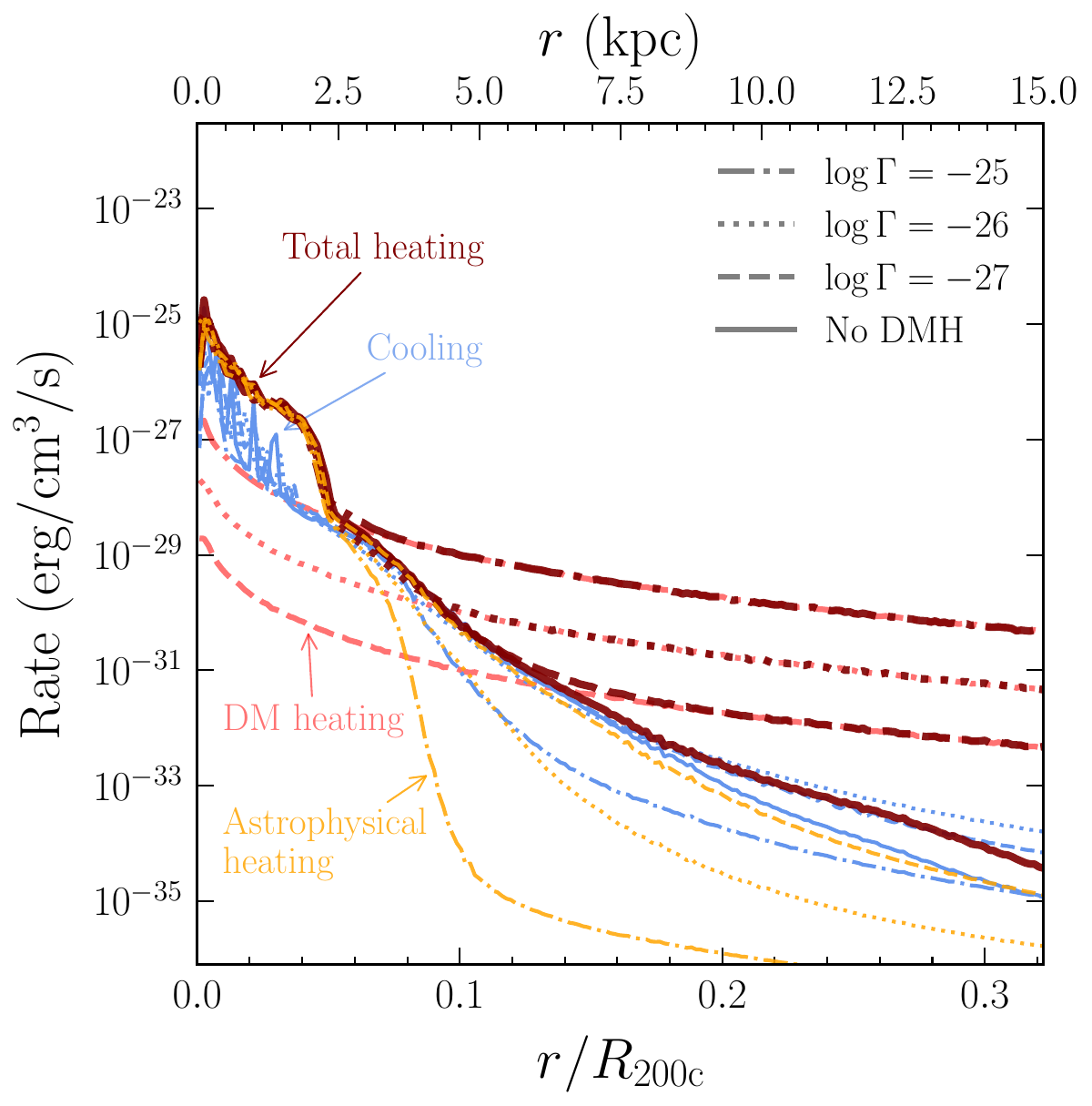}
\caption{Volume-averaged radial profiles of the heating and cooling rates in the WLM-like dwarf simulations at $t=1.5$~Gyr. Line style identifies the effective DM heat-transfer rate $\Gamma_\mathrm{heat}$ from Equation~\ref{eq:DM_heating_rate}, including the control run without DM heating. The total heating rate (dark red) is decomposed into astrophysical (yellow) and DM (light red) contributions; the total cooling rate is blue. Unlike astrophysical heating, DM heating declines gradually with radius and dominates the heating budget in the CGM.} \label{fig:radial_profiles}
\end{figure}

\subsection{Heating and Cooling Rate Calculations}
\label{sec:Cooling_Rates}

For the diagnostics in Section~\ref{sec:results} we evaluate the volumetric
heating and cooling rates as RAMSES does internally. The cooling rate is the sum
of a metal-free term and a metal-line term scaled by metallicity,

\begin{equation}
    C\,[{\rm erg\,cm^{-3}\,s^{-1}}] = \left[\Lambda_{\rm H,He}(n_{\rm H},T)
      + \frac{Z}{Z_\odot}\,\Lambda_{Z}(n_{\rm H},T)\right] n_{\rm H}^2 ,
\label{eq:AstroCoolingRate}
\end{equation}
where $\Lambda_{\rm H,He}$ is the metal-free cooling function \citep{Katz1996}, including bremsstrahlung, collisional ionization, recombination, and line cooling, and $\Lambda_{Z}$ follows \citet{Sutherland1993}. Both are read from the cooling tables written with each snapshot and interpolated at each cell's density and temperature. Only the metal term carries a metallicity dependence.

The astrophysical heating rate has two contributions: photoionization heating of H and He by the metagalactic UV background and supernova feedback. For the latter, the momentum-based feedback scheme injects energy directly into gas cells rather than through the cooling function; we therefore reconstruct it from the star particles using the injection model described in Section~\ref{subsec:simphysics}.

\begin{figure*}
\centering
\includegraphics[width=\linewidth]{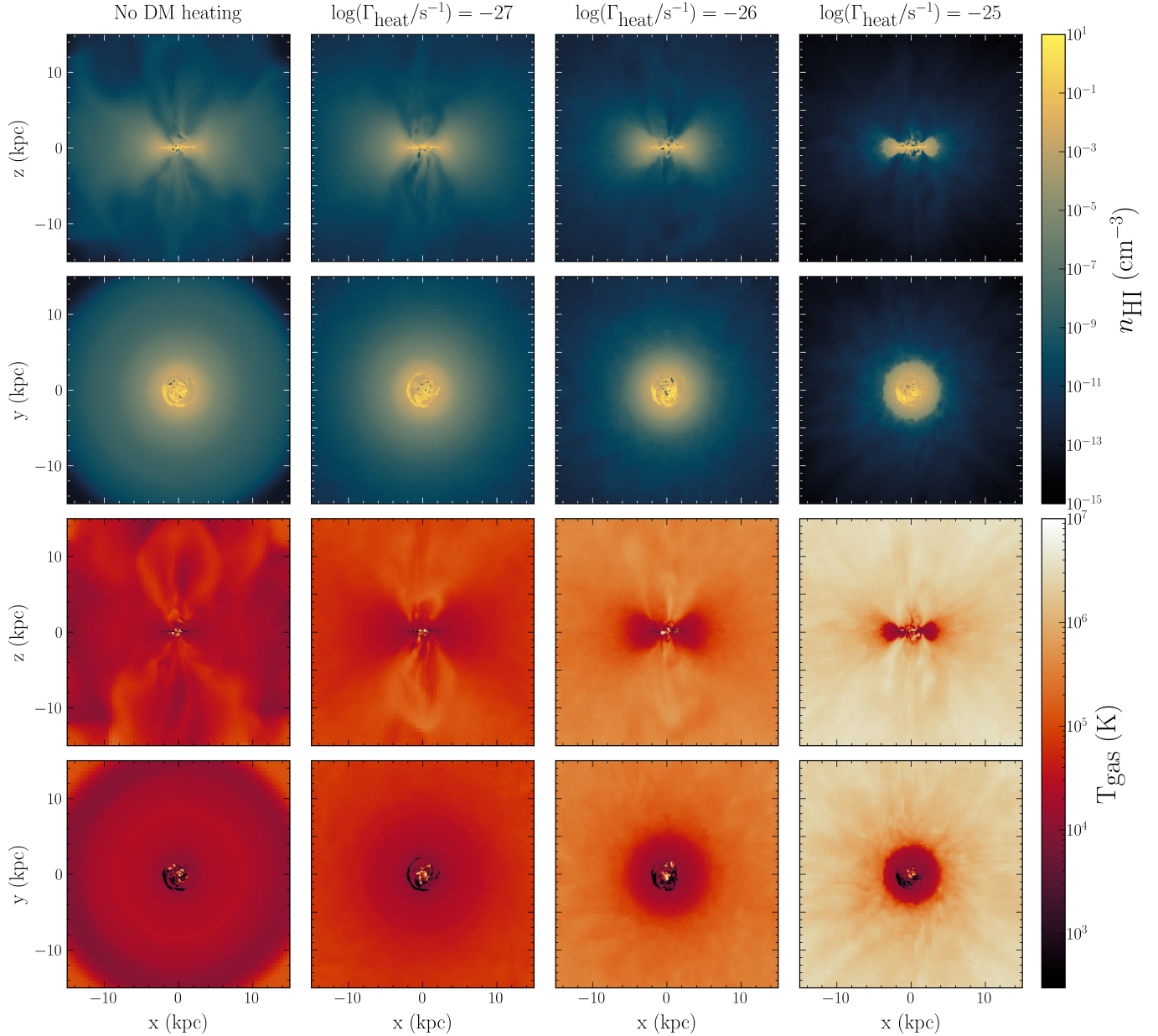}
\caption{Neutral-hydrogen number density (top two rows) and gas temperature (bottom two rows) at $t=1.5$~Gyr, shown in edge-on and face-on slices through the WLM-like galaxy. Columns show the control simulation and increasing values of the effective DM heat-transfer rate $\Gamma_\mathrm{heat}$. Larger $\Gamma_\textrm{heat}$ produces a hotter, more diffuse CGM while leaving the dense disk comparatively unchanged.}
\label{fig:slice_rho_T}
\end{figure*}

\subsection{Synthetic Column Densities and Radial Selection}
\label{sec:synthetic_column_densities}

To compute synthetic column densities from the RAMSES simulations, we post-process each simulation snapshot using the publicly available ionization tables of \citet{Plo20_MetalIon}. These tables were generated with the photoionization code Cloudy \citep{cloudy} over a wide range of gas density, temperature, metallicity, and redshift, while accounting for a metagalactic UV/X-ray background, dust, cosmic rays, local radiation fields, and self-shielding. For each simulation cell, we use the gas density, temperature, metallicity, and redshift to interpolate the tabulated ion fractions, $f_{\rm ion}(n_{\rm H},T,Z,z)$. The number density of a given ion is then computed as $n_{\rm ion}=f_{\rm ion}(n_{\rm H},T,Z,z) A_X (Z/Z_\odot)n_{\rm H}$, where $A_X$ is the solar abundance of the corresponding element. We construct projected column-density maps by integrating the ion number densities along the line of sight. This procedure allows us to predict the distributions of \ion{H}{1}, \ion{C}{2}, \ion{C}{4}, \ion{Si}{2}, \ion{Si}{3}, and \ion{Si}{4} without performing additional Cloudy calculations for each simulation output.

A limitation of our simulations is that they use isolated rather than cosmological initial conditions. In cosmological simulations, the CGM is continuously replenished and reshaped by large-scale structure through filamentary accretion, infalling satellites, recycled galactic winds, and the ongoing growth of the dark matter halo. These processes add gas, metals, and thermal energy to the CGM and can modify its density and ionization structure, especially in the outer halo. In contrast, our isolated setup initializes the halo within a low-density ambient medium and does not include subsequent cosmological inflows. As a result, the outer CGM near the virial radius may not accurately reproduce the gas distribution expected in a fully cosmological environment. We therefore restrict the column-density comparison throughout this paper to projected radii $d_{\rm proj}<15\,{\rm kpc}$, or equivalently $d_{\rm proj}/R_{\rm 200c}\lesssim0.3$ for the WLM-like simulation. Within this region, the mean enclosed hydrogen density exceeds 200 times the cosmic mean density, indicating that the gas is dominated by material associated with the collapsed halo rather than by the imposed background medium.

\begin{figure*}[th]
    \centering
    \includegraphics[width=\linewidth]{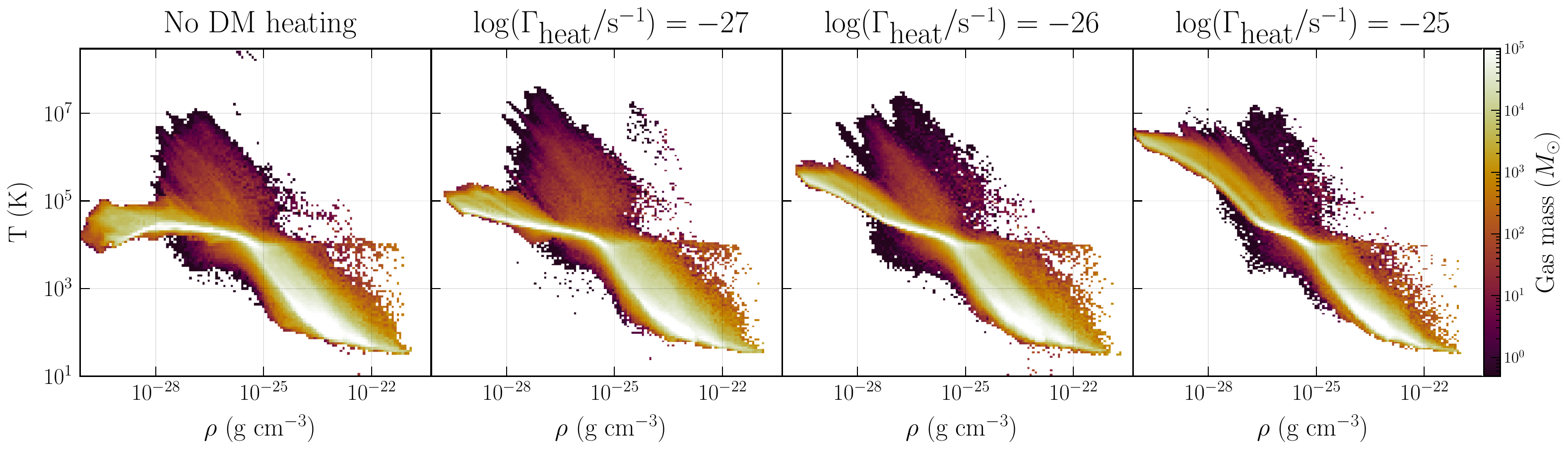}
    \caption{Gas temperature versus mass density at $t=1.5$~Gyr for the control simulation and increasing values of the effective DM heat-transfer rate $\Gamma_\mathrm{heat}$. Color indicates the gas mass in each phase-space bin. DM heating shifts the low-density CGM gas to higher temperatures, while the high-density ISM remains comparatively unchanged.}
    \label{fig:phase_space}
    \end{figure*}

\section{Results}
\label{sec:results}

\subsection{Heating and Cooling Rates}

We plot the heating and cooling rates from RAMSES as a binned function of radius in \autoref{fig:radial_profiles}. Because the volumetric photoheating rate tracks the neutral hydrogen density, it falls off dramatically between the disk and the outer CGM, whereas the DM heating rate follows $\rho_{\rm DM}$ and declines much more gradually. Astrophysical heating dominates near the galaxy's disk, while DM heating becomes the dominant heating term in the CGM for all three values of $\Gamma_{\rm heat}$ considered here. The cooling rate drops monotonically with radius beyond the inner disk as the gas density declines.

\subsection{DM Heating Has Minimal Effect on the ISM but Significantly Affects the CGM}

Including non-standard heating due to DM significantly changes the extended CGM, whereas the neutral gas in the disk is largely unaffected. \autoref{fig:slice_rho_T} shows neutral-hydrogen density and gas-temperature slices through the simulated CGM and ISM. The largest change occurs in the CGM, while the ISM remains close to the control run. \autoref{fig:phase_space} shows the same trend in gas phase space: low-density CGM gas is most strongly affected by DM heating, while the high-density ISM remains nearly unchanged.

This contrast is also evident in the integrated properties of the disk. Within the disk (the cylinder defined by $r<5$~kpc and $|z|<2$~kpc), the star formation rate averaged over the final 100~Myr is $2.2$--$2.3\times10^{-3}\,M_\odot\,\mathrm{yr}^{-1}$ in all four runs, varying by less than 5\% and showing no trend with $\Gamma_{\rm heat}$. The disk neutral-hydrogen mass declines monotonically with heating rate, by 3\%, 9\%, and 21\% for $\Gamma_{\rm heat}=10^{-27}$, $10^{-26}$, and $10^{-25}$~\persec, respectively. Both changes are modest compared with the CGM response described below, where the neutral-hydrogen column density falls by more than 3~dex over the same range. That the star formation rate is unchanged while the CGM columns drop by orders of magnitude indicates that the effect is not mediated by a change in
stellar feedback.

\begin{figure*}[th]
\centering
\includegraphics[width=\linewidth]{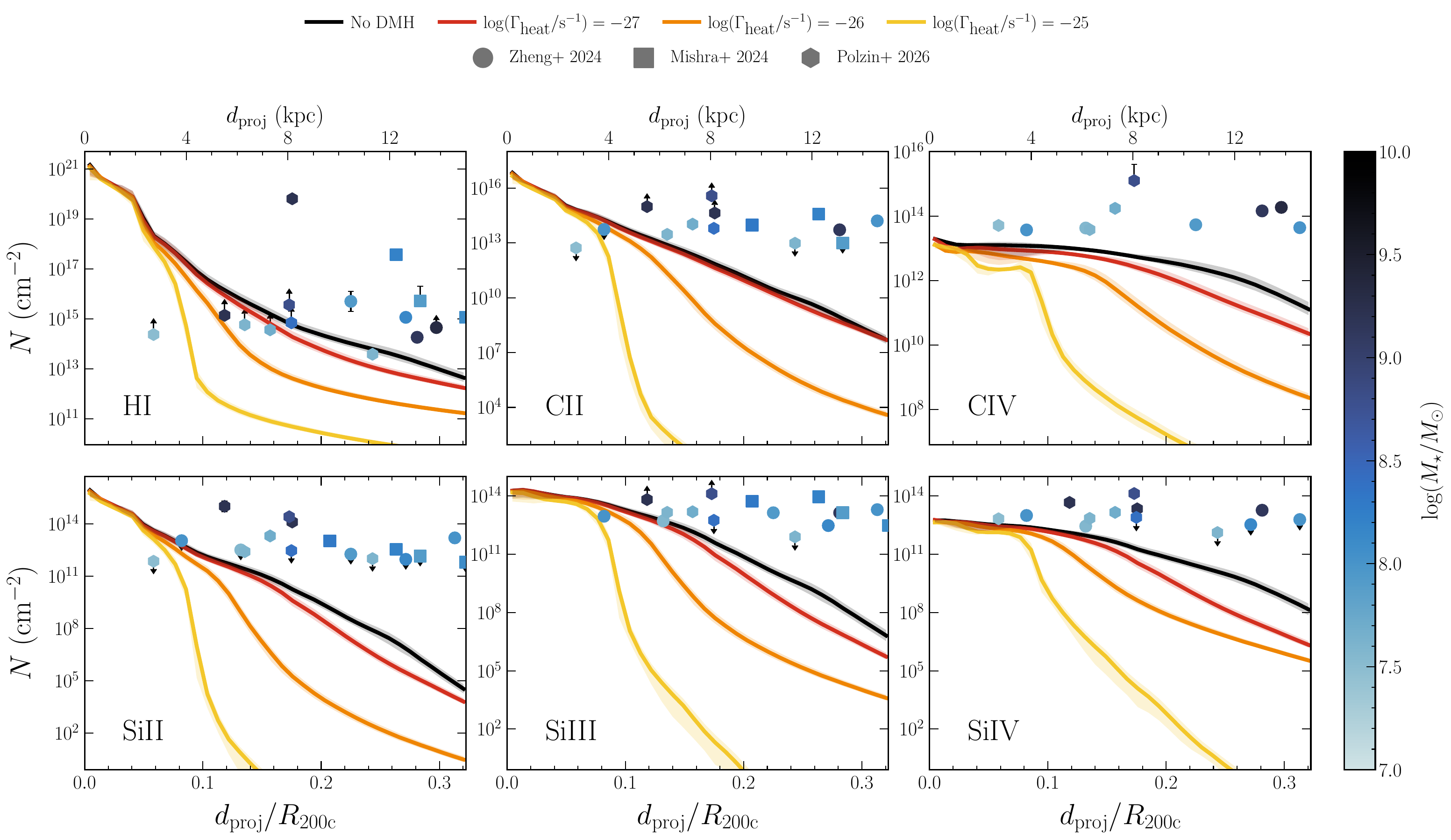}
\caption{Projected column-density profiles of \ion{H}{1}, \ion{C}{2}, \ion{C}{4}, \ion{Si}{2}, \ion{Si}{3}, and \ion{Si}{4} around the simulated WLM-like dwarf galaxy at $t=1.5$~Gyr. The black curve in each panel shows the control simulation without DM heating, while the colored curves show $\Gamma_{\rm heat}=10^{-27}$, $10^{-26}$, and $10^{-25}$~\persec. Quasar absorption-line measurements compiled by \citet{Zhe24_CGM_observations}, \citet{Mis24_CGM_observations}, and \citet{Polzin2026} are shown as circles, squares, and hexagons, respectively; marker color denotes the host-galaxy stellar mass, and arrows indicate upper or lower limits. The lower axes give projected radius normalized by $R_{\rm 200c}$ for the observations and simulations, while the upper axes give physical distance for the simulated galaxy. Because the simulated galaxy has $M_\star\approx10^7\,M_\odot$, below most of the observed hosts, the observational comparison is illustrative rather than matched. Increasing $\Gamma_{\rm heat}$ suppresses every column-density profile, with a stronger effect on low-ionization species than on high-ionization species. Simulations matched to the observed galaxy population could therefore enable CGM absorption measurements to test DM-heating models.} 
\label{fig:metal_profs}
\end{figure*}

\subsection{Neutral-hydrogen and Metal-ion Column-density Signatures}

As described in Section~\ref{sec:synthetic_column_densities}, we construct the column density of each ion by interpolating its ionization fraction from the density, temperature, metallicity, and redshift of every simulation cell, and then integrating the resulting ion number density along the line of sight. Figure~\ref{fig:metal_profs} shows the projected radial profiles for the four simulations. DM heating suppresses the neutral-hydrogen column throughout the CGM. At $d_{\rm proj}=10$~kpc, the suppression is 0.6, 2.1, and 3.9~dex for $\Gamma_{\rm heat}=10^{-27}$, $10^{-26}$, and $10^{-25}$~\persec, respectively.

The metal ions respond more strongly than \ion{H}{1}, and the low-ionization states respond more strongly than the high-ionization states. At 10~kpc and $\Gamma_{\rm heat}=10^{-26}$~\persec, the columns fall by 5.5~dex for \ion{Si}{2}, 4.2~dex for \ion{Si}{3}, 4.1~dex for \ion{C}{2}, 3.0~dex for \ion{Si}{4}, and 2.2~dex for \ion{C}{4}, compared with 2.1~dex for \ion{H}{1}.

The magnitude of these changes motivates comparison with CGM column-density observations of nearby dwarfs. Figure~\ref{fig:metal_profs} includes the compilation of \citet{Zhe24_CGM_observations}, which incorporates measurements from \citet{2014ApJ...796..136B}, \citet{2018ApJ...864..132B}, \citet{2014MNRAS.445.2061L}, and \citet{Joh17_CGM_observations}, among others, as well as data from \citet{Mis24_CGM_observations} and \citet{Polzin2026}. Each of the three samples uses a different definition of the virial radius and a different cosmology, so we standardize them to our conventions as described in \autoref{app:radii}.

Although the observed samples extend farther, we restrict the comparison to $d_\text{proj}/R_{\rm 200c}\lesssim 0.3$, because our isolated simulations are not designed to model the outer halo and surrounding environment at larger radii (Section~\ref{sec:synthetic_column_densities}). Within this range, the control simulation overlaps some of the lower observed columns at small projected radii but falls below most of the metal-ion measurements, with the disagreement generally increasing toward larger radii. This comparison is necessarily qualitative because our simulated galaxy lies near the low-mass end of the observational sample. More massive halos are expected to retain larger columns of CGM gas and metals \citep{Zhe24_CGM_observations, Pia25_Dwarf_CGM_sims}, and differences in their enrichment and feedback histories can further affect the metal-ion columns. In addition, our isolated simulation lacks cosmological accretion, recycled winds, and infalling satellites, which can replenish the outer CGM. These effects plausibly contribute to the discrepancy (and its increasing trend with radius).

A robust observational comparison requires a matched cosmological simulation suite, as discussed in Section~\ref{sec:cosmological_simulations}. We therefore interpret the systematic, ion-dependent changes relative to the control run as the principal result of Figure~\ref{fig:metal_profs}, rather than using the absolute agreement with the heterogeneous observations to constrain $\Gamma_{\rm heat}$.

\section{Discussion and Conclusions}
\label{sec:discussion_conclusions}

We have implemented a local, model-independent DM heating prescription in RAMSES and applied it to an isolated, WLM-like dwarf galaxy. The simulations show that the CGM is more sensitive to DM heating than the dense ISM, making CGM column densities a promising observable for constraining anomalous heat transfer from dark matter.

\subsection{Relation to Previous Leo T Calorimetric Constraints}
\label{sec:leoT_constraints}

The approach developed here is complementary to previous calorimetric studies of gas-rich dwarfs such as Leo T \citep{WadWan22}. Those calculations assumed that the gas is in hydrostatic equilibrium (see \citet{faerman13}) and required anomalous DM heating not to exceed the astrophysical cooling rate. Here, we instead use hydrodynamic simulations to follow how DM heating modifies the gas distribution, temperature, and observable column densities in an isolated dwarf galaxy. This simulation-based approach is less analytically transparent than the Leo T cooling argument, but it allows ordinary astrophysical heating, cooling, feedback, and gas motion to enter the calculation self-consistently.

For reference, these Leo T calorimetric limits can be recast in terms of the effective heating parameter used here. Using $\Gamma_{\rm heat}=\Gamma_{\rm decay}E_{\rm heat}/E_{\rm particle}$, they imply approximately $\Gamma_{\rm heat}\lesssim10^{-26}\ {\rm s}^{-1}$ for DM decay to $e^+e^-$ over $m_{\rm DM}=10^6$--$10^7$~eV, and a similar bound for decay to $\gamma\gamma$ over $m_{\rm DM}=20$--$200$~eV. These benchmarks show that the range simulated here brackets effective heating rates already probed using Leo T, but they are not constraints derived from the WLM-like simulations presented in this work.

\subsection{Astrophysical Systematics and Observational Interpretation}
\label{sec:astrophysical_systematics}

Other dwarf-CGM simulations have shown that \ion{C}{4} strongly depends on galaxy stellar mass \citep{Pia25_Dwarf_CGM_sims}. A similar trend is reported for EAGLE halos in Figure~8 of \citet{Zhe24_CGM_observations}, where the \ion{H}{1} and metal-ion column densities increase with halo mass. This dependence may explain part of the disagreement at larger radii between the smaller WLM-like galaxy in our simulation and the observations shown in \autoref{fig:metal_profs} for larger galaxies.

Recent observations have identified a dearth of low- and intermediate-ionization metal absorption relative to high-ionization states \citep{Joh17_CGM_observations, Zhe24_CGM_observations, Mis24_CGM_observations}. Although this pattern could arise from purely astrophysical effects such as low gas density and a high volume-filling factor, our results suggest that non-standard DM heating can shift the ionization balance in the same direction. Because the CGM expands as it is heated, all metal-ion columns are suppressed in the simulations with DM heating, but the low ions are suppressed considerably more than the high ions. At $\Gamma_{\rm heat}=10^{-26}$~\persec\ and $d_{\rm proj}=10$~kpc, \ion{C}{2} falls by 4.1~dex while \ion{C}{4} falls by 2.2~dex, and \ion{Si}{2} falls by 5.5~dex while \ion{Si}{4} falls by 3.0~dex; the \ion{C}{2}/\ion{C}{4} and \ion{Si}{2}/\ion{Si}{4} ratios therefore drop by 1.8 and 2.4~dex, respectively. A quantitative interpretation comparing our DM-heating models with astrophysical data would require marginalizing over ordinary
astrophysical uncertainties.

\subsection{Mapping the Heating Rate to Particle Models}
\label{sec:mapping_to_particle_models}

We parameterize DM heating through the effective local heat-transfer rate $\Gamma_{\rm heat}$ rather than committing to a particular particle model. The implementation used in this paper corresponds to a decay-like local source term,
\begin{equation}
    \dot{u}_{\rm heat} = \rho_{\rm DM}\, c^2\,\Gamma_{\rm heat},
\end{equation}
where $\dot{u}_{\rm heat}$ is the injected thermal energy per unit volume per unit time. Any bounds on $\Gamma_{\rm heat}$ inferred from CGM column densities can be mapped to particular DM models, such as axion-like particles, sterile neutrinos, primordial black holes, dark photon DM, or excited DM, by accounting for the fraction of injected energy deposited as local heat, as illustrated by the Leo T benchmarks in Section~\ref{sec:leoT_constraints}. For low-energy electrons and photons, the deposited heat depends on how the injected energy is partitioned among heating, ionization, and excitation of the gas \citep{Shu85,Ric02,Fur10}.

Decay is not the only particle-physics channel that can source gas heating. Other scenarios predict different local density scalings and can be incorporated by replacing the source term above with the appropriate model-dependent expression. For example, annihilation-like heating scales as $\dot{u}_{\rm heat}\propto \rho_{\rm DM}^2$, with the proportionality constant set by the annihilation cross section, the DM mass, and the fraction of the annihilation energy deposited as heat. Scattering or drag-mediated heating instead depends on both the DM density and the density of the target gas species, giving source terms of the schematic form $\dot{u}_{\rm heat}\propto \rho_{\rm DM} n_{\rm H}$ or $\dot{u}_{\rm heat}\propto \rho_{\rm DM} n_e$ for interactions with neutral hydrogen or free electrons. These channels are therefore more centrally weighted than the decay-like source considered here, especially for annihilation, and a dedicated simulation would be needed to translate our CGM constraints to those models quantitatively. The numerical framework is general, however: the same RAMSES patch can be extended by replacing the local heating-rate prescription with the desired dependence on $\rho_{\rm DM}$, $n_{\rm H}$, $n_e$, temperature, or velocity. Appendix~\ref{app:timestep} connects these source dependencies to their timestep treatment, distinguishing concentrated but gas-state-independent sources from interactions that can become stiff because they depend on the gas state.

\subsection{Limitations and Future Directions}
\label{sec:cosmological_simulations}

We have only considered local energy injection in this paper. Non-local energy injection could be implemented following approaches similar to \citet{Sch14, Sch18_DM_annihilation}. Our simulations include a metagalactic UV photoionizing background and a
density-dependent self-shielding prescription, but not full radiative transfer.

The simulations presented here are isolated rather than cosmological. Cosmological simulations indicate that the CGM is generally a mixture of intergalactic accretion, recycled winds from the central galaxy, and material associated with infalling satellites, highlighting the importance of these processes for the structure of the outer halo. Extending the present study to fully cosmological simulations is therefore an important direction for future work. A robust comparison with the heterogeneous observations will require simulations spanning the stellar and halo masses of the observed galaxies, their enrichment and feedback histories, and the DM-heating rates considered here. A likelihood analysis can then compare synthetic and observed CGM column densities, including the measurements of \citet{Zhe24_CGM_observations}, \citet{Mis24_CGM_observations}, and \citet{Polzin2026} out to the virial radius. We leave this analysis to future work.

We also plan to extend our analysis from WLM to smaller gas-rich galaxies such as Leo T and Leo P, and to compact neutral-hydrogen clouds such as Cloud-9. Existing RAMSES simulations of Leo T provide a useful starting point for such extensions \citep{Blana2024_LeoT_RAMSES}. We expect DM heating effects to be stronger in these systems because their gas cooling rates are lower than in WLM \citep{WadFar21}. Although no quasar sightlines have been observed through the CGM of Leo T, simulations of its ISM could still provide useful constraints on DM heating.

Upcoming optical and 21~cm surveys will have unprecedented detection sensitivity, especially for isolated dwarfs \citep{LSST,Drl19_LSST,WFIRST,Aih18,DESI,SKA,Kor20_Wallaby,Mad21,van22,Zha21,Mut21}. These surveys could substantially increase the sample of isolated dwarf galaxies similar to WLM and Leo T. A larger sample observed along quasar sightlines would enable searches for CGM heating signatures across galaxy mass, environment, and star-formation history. Together with cosmological simulations, such observations could turn the pronounced, ion-dependent CGM response found here into quantitative constraints on heat exchange between DM and ordinary matter.

\begin{acknowledgements}
We thank Sandip Roy, Marc Kamionkowski, Glennys Farrar, Tracy Slatyer, Hongwan Liu, and Wenzer Qin for useful discussions.

A.M. acknowledges support from the National Science Foundation Graduate Research Fellowship under Grant No. 2039656.
\end{acknowledgements}

\section*{Data and Code Availability}
The DM-heating implementation for \texttt{RAMSES} is publicly available at the GitHub repository listed in Footnote~\ref{fn:code}.

\appendix
\section{Dependence on Simulation Setup}\label{sec:resolution_uv_dependence}

\begin{figure*}[t]
\centering
\includegraphics[width=\linewidth]{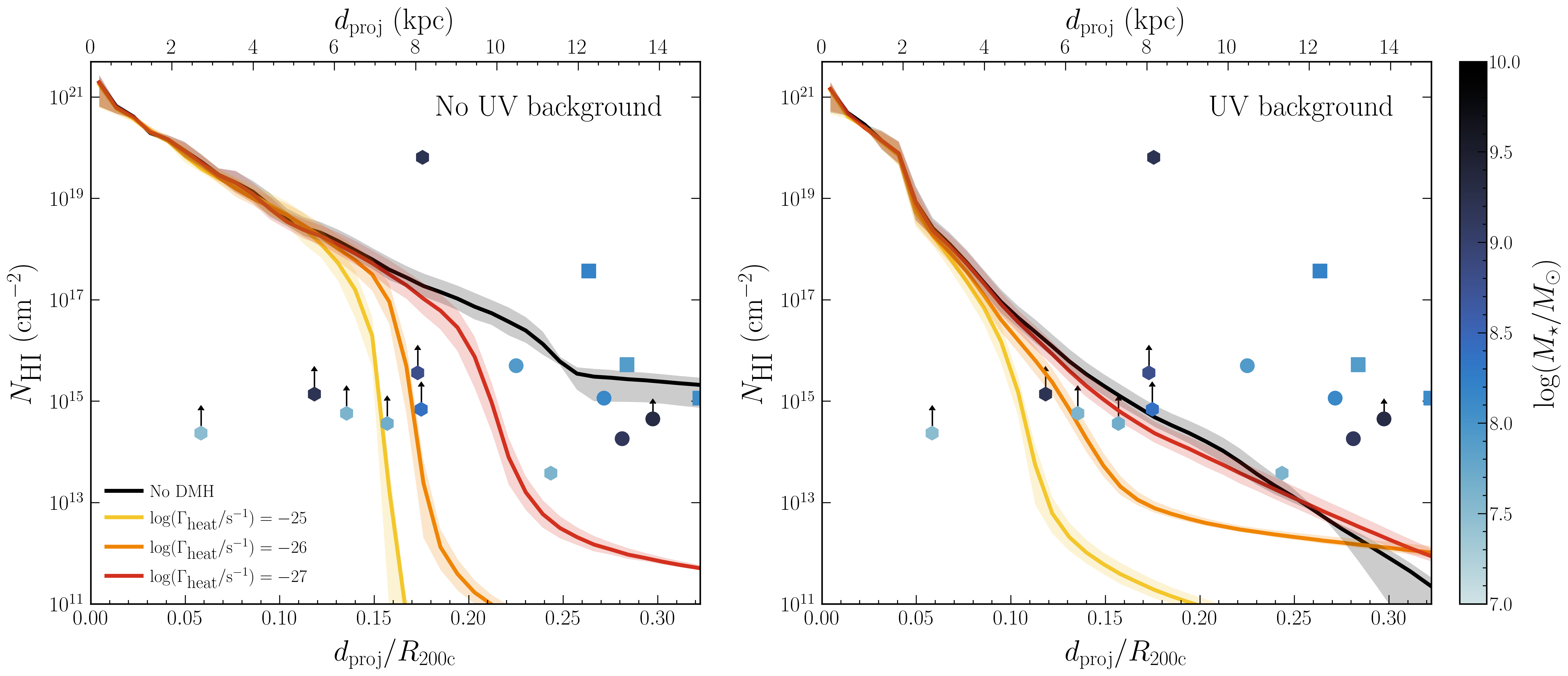}
\caption{Projected neutral-hydrogen column-density profiles at $t=0.7$~Gyr. The right panel is similar to the upper-left panel of Figure~\ref{fig:metal_profs} with the metagalactic UV background but at a different timestep, while the left panel shows otherwise identical simulations without it. Plotting conventions and observational markers follow Figure~\ref{fig:metal_profs}. The UV background substantially suppresses the neutral-hydrogen column, but increasing $\Gamma_{\rm heat}$ produces a strong additional suppression in both cases.}
\label{fig:resolution_uv_dependence}
\end{figure*}

Figure~\ref{fig:resolution_uv_dependence} shows the effect of the metagalactic UV
background on the CGM column densities at $l_{\max}=14$ and $t=0.7$ Gyr: it ionizes the gas and substantially reduces the neutral-hydrogen column. 

Numerical convergence in this problem requires two separate tests, because the adaptive mesh resolves the disk and the CGM by different mechanisms. RAMSES
refines following a quasi-Lagrangian scheme, triggering on enclosed mass, so refinement is concentrated in dense gas: in our fiducial runs it is confined to a $\sim10$~kpc region around the disk, and $\sim60$\% of all cells sit at the base grid. Raising $l_{\max}$ therefore buys resolution only where the gas is already dense, while the diffuse CGM that our column densities probe remains at the coarsest cell size regardless.

We first vary $l_{\max}$. In addition to the fiducial runs with $l_{\max}=14$ (minimum cell size $\Delta x = L_{\rm box}/2^{l_{\max}}\simeq4$~pc), we performed lower-resolution simulations with $l_{\max}=12$ ($\simeq15$~pc), integrated up to 0.7 Gyr. The gas density, temperature, and ion column-density profiles are qualitatively unchanged between the two. Because the two share $l_{\min}$ and hence an identical grid throughout the CGM, this test constrains the disk rather than the region our column densities probe. 

We therefore ran a second set in which the base grid itself is refined, with $l_{\min}=8$ rather than 7, halving the cell size everywhere outside the refined region from 469 to 234~pc. These runs use identical initial conditions, refinement criteria, and star formation threshold, so the difference isolates the effect of resolution in the CGM. The resulting column densities agree with the fiducial runs at $t=0.7$ Gyr to within 0.09~dex at radii below 15~kpc. At 15~kpc, the control run differs by up to 0.17~dex while the heated runs remain within 0.05~dex; the finer base grid yields slightly higher values. The disk neutral-hydrogen mass agrees to within 2\% throughout. These offsets are an order of magnitude smaller than the DM-heating signal at the same radii, and we conclude that the CGM column densities are converged with respect to both the peak and the base resolution. We present the $l_{\max}=14$, $l_{\min}=7$ simulations in the main text.

These grid-resolution tests do not test convergence with respect to the dark matter particle sampling. Because $\rho_{\rm DM}$ is reconstructed by depositing a finite number of particles onto the mesh, it contains shot noise set by the number of particles contributing to each cell. The cloud-in-cell scheme used here smooths this noise across neighboring cells, but the noise can become more important in the outer halo as the dark matter particle density decreases. Since the heating rate in Equation~\ref{eq:DM_heating_rate} is proportional to $\rho_{\rm DM}$, fluctuations in the reconstructed density directly produce fluctuations in the local heating rate. We have not varied the dark matter particle number or the deposition scheme, so convergence with respect to this effect remains to be established. Future tests could increase the particle number or use a higher-order deposition scheme, such as triangular-shaped-cloud deposition, and quantify the resulting changes in the CGM column densities.

\section{Operator Splitting and Timestep Control}
\label{app:timestep}

Our implementation advances hydrodynamics, DM heating, and radiative cooling as successive updates rather than solving for all three processes simultaneously. This procedure is known as \emph{operator splitting}. Specifically, \texttt{RAMSES} first performs the hydrodynamic update, then adds the energy in Equation~\ref{eq:update}, and finally applies cooling and chemistry. Operator splitting separates two numerical questions: whether each update is stable and whether applying the updates sequentially is an accurate approximation to their coupled evolution.

\subsection{Stability of the Heating Update}

Radiative cooling can impose a stability constraint because it removes energy at a rate that depends strongly on the gas state. If the cooling time becomes shorter than the timestep, a simple explicit update can remove more energy than the cell contains. Cooling algorithms must therefore limit the timestep or integrate the loss term implicitly.

The DM-heating update is simpler. For the decay-like source considered here, the deposited energy is positive and independent of the gas internal energy. Holding the heating rate fixed during a timestep gives $\Delta E_{\rm gas}=\dot{u}_{\rm heat}\Delta t$, so the update is exact and cannot make the internal energy negative. The source therefore requires no stability limiter. This property is shared by particle- and grid-based implementations, as well as by an annihilation-like source with a fixed deposited rate. The advantage of the grid-based implementation instead comes from its smooth spatial deposition and synchronized cell updates, as discussed in Section~\ref{sec:timestep}.

\subsection{Accuracy of the Split Update}

Stability of the heating update does not by itself guarantee that the combined evolution is accurate. Because the hydrodynamic fluxes are computed before the heat is added, they do not respond to the pressure increase until the next update. Cooling is likewise applied after heating rather than simultaneously with it. This ordering is accurate when the heating changes the gas energy by only a small fraction during one timestep. We quantify that change by
\begin{equation}
\chi \equiv
\frac{\rho_{\rm DM}\, c^{2}\, \Gamma_{\rm heat}\, \Delta t}
     {\rho_{\rm gas}\, \varepsilon},
\qquad
\varepsilon = \frac{3}{2}\frac{k_{\rm B} T}{\mu m_{p}},
\label{eq:chi}
\end{equation}
where $\varepsilon$ is the gas specific internal energy, $\mu$ is the mean molecular weight, and $m_p$ is the proton mass. Equivalently, $\chi$ is the ratio of the timestep to the local heating time, $t_{\rm heat}=\rho_{\rm gas}\varepsilon/(\rho_{\rm DM}c^2\Gamma_{\rm heat})$. The split update is accurate when $\chi\ll1$. For our decay-like source, $\chi \propto (\rho_{\rm DM}/\rho_{\rm gas})\,T^{-1}$ and therefore tends to be largest in tenuous, cool gas, where a fixed amount of deposited energy produces a larger fractional change in the gas internal energy.

For sufficiently large $\Gamma_{\rm heat}$, Equation~\ref{eq:chi} provides a direct prescription for controlling the splitting error. Requiring $\chi<f$, where $f<1$ is the chosen maximum fractional energy change per step, gives the additional timestep bound
\begin{equation}
\Delta t_{\rm heat}
= f\frac{\rho_{\rm gas}\varepsilon}{\dot{u}_{\rm heat}}
= f\frac{\rho_{\rm gas}\varepsilon}
{\rho_{\rm DM}c^2\Gamma_{\rm heat}}.
\label{eq:heating_timestep}
\end{equation}
The simulation timestep can then be chosen as the smaller of the standard \texttt{RAMSES} timestep and $\Delta t_{\rm heat}$. This is an accuracy requirement, not a stability requirement: it ensures that the pressure changes gradually enough for the hydrodynamic and cooling updates to follow the added heat. When evaluating Equation~\ref{eq:heating_timestep}, $\varepsilon$ should be bounded below by the value at the cooling temperature floor so that spuriously cold cells do not impose arbitrarily short timesteps.

The same prescription extends to the alternative heating channels introduced in Section~\ref{sec:mapping_to_particle_models} by using their model-dependent volumetric heating rate $\dot{u}_{\rm heat}$. An annihilation-like source remains positive and independent of the gas internal energy, but its $\rho_{\rm DM}^{2}$ scaling concentrates the heating toward the halo center and can make the accuracy bound more restrictive.
By contrast, a scattering or thermal-contact source can depend on the gas state. A source that relaxes the gas temperature toward the effective DM temperature $T_{\chi}$ on a timescale $\tau$ is genuinely stiff when $\tau < \Delta t$ and should be advanced with the exponential update $T \rightarrow T_{\chi} + (T - T_{\chi})\,e^{-\Delta t/\tau}$, which is unconditionally stable.

\section{Standardizing Radii for Comparison}
\label{app:radii}

The simulation and the three observational samples we compare against \citep{Zhe24_CGM_observations, Mis24_CGM_observations, Polzin2026} each use a different virial-radius definition, so their normalized impact parameters are not directly comparable. We standardize the three observational samples to $R_{\rm 200c}$, the radius enclosing a mean density of $200\rho_{\rm crit}(z)$ in our adopted cosmology; this radius normalizes the horizontal axis of Figure~\ref{fig:metal_profs}. The simulated halo's $R_{\rm 200c}$ is likewise evaluated in this cosmology rather than in the cosmology assumed internally by \texttt{DICE}.

The conversion is not simply $R_2/R_1=(\Delta_1/\Delta_2)^{1/3}$, where $\Delta_1$ and $\Delta_2$ are the overdensities defining the original and target radii. That scaling holds only if all the enclosed mass lies inside $R_1$, whereas an NFW halo continues to accumulate mass outward: less mass is enclosed at the smaller radius than the fixed-mass scaling assumes, so the true radius is smaller still. For the \citet{Zhe24_CGM_observations} conversion, for example, the uniform-sphere estimate gives 0.677 while the NFW correction is 0.612, a 10\% difference.

We instead assume an NFW profile. Each quoted radius comes with the mean density it encloses by definition ($200\rho_{\rm m}$ for $R_{\rm 200m}$, $\Delta_c(z)\rho_{\rm crit}(z)$ for $r_{\rm vir}$) and that density, the radius, and an assumed concentration together fix the profile completely: the scale radius is $r_s=R/c$, and the normalization follows from requiring the enclosed NFW mass to reproduce the stated mean density. We then solve numerically for the radius at which the enclosed mean density equals $200\rho_{\rm crit}$. For the observed galaxies we adopt a concentration $c=10$, typical of halos in
this mass range \citep[e.g.][]{Dutton2014}.

\citet{Zhe24_CGM_observations} quote $R_{\rm 200m}$, defined against $200\rho_{\rm m}$ rather than $200\rho_{\rm crit}$. Both are evaluated at $z=0$, so for every galaxy in the sample the two thresholds differ by the same constant factor $\rho_{\rm m}/\rho_{\rm crit}=\Omega_{\rm m}=0.310$ \citep{Planck2020}. The NFW normalization and scale radius cancel in the conversion, which therefore depends only on the two thresholds and the assumed concentration rather than on the halo's mass or size, and a single factor applies to the whole sample: $R_{\rm 200c}=0.612\,R_{\rm 200m}$.

\citet{Mis24_CGM_observations} quote $R_{\rm vir}$ defined by the \citet{Bryan1998} overdensity $\Delta_c(z)\rho_{\rm crit}(z)$, with $\Delta_c=18\pi^2+82x-39x^2$ and $x=\Omega_{\rm m}(z)-1$, evaluated in their cosmology ($H_0=70$, $\Omega_{\rm m}=0.3$), for which $\Delta_c=101$ at $z=0$. Because their $\rho_{\rm crit}$ differs from ours, it does not cancel between the two thresholds. The correction is redshift dependent, running from $R_{\rm 200c}=0.796\,R_{\rm vir}$ at $z=0.08$ to $0.902\,R_{\rm vir}$ at $z=0.72$, and we therefore apply it per galaxy rather than using a sample median.

\citet{Polzin2026} quote $R_{\rm 200c}$ for $H_0=70$~km~s$^{-1}$~Mpc$^{-1}$. The overdensity and reference density already match ours and only $H_0$ differs, making this the smallest correction: we multiply their quoted radii by 1.028.

\clearpage

\bibliography{DM_heating,leoT_extra_refs}
\bibliographystyle{aasjournalv7}

\end{document}